\documentclass[journal]{IEEEtran}

\usepackage[mode=image]{standalone} 

\usepackage[dvipsnames]{xcolor}
\usepackage{soul,framed} 
\usepackage{fancyhdr}

\colorlet{shadecolor}{yellow}
\usepackage{graphicx}
\usepackage{amsfonts}
\usepackage{multirow}
\usepackage{multicol}
\usepackage{makecell}

\usepackage[cmex10]{amsmath}
\usepackage{amssymb}
\usepackage{array}
\usepackage{mdwmath}
\usepackage{mdwtab}
\usepackage{eqparbox}
\usepackage{url,balance,comment}
\usepackage{siunitx}
\usepackage[normalem]{ulem}

\usepackage{tikz}
\usepackage{tikzscale}
\usepackage{xifthen}
\usepackage{pgfplots}
\usepackage{pgfplotstable}
\pgfplotsset{compat = newest}

\usepgfplotslibrary{groupplots}

\usepackage{graphicx}

\usepackage{bm}
\usepackage{enumerate}
\usepackage{tkz-euclide} 
\usetikzlibrary{arrows.meta}

\usetikzlibrary{calc,positioning,shapes,fit}

\tikzset{>=latex}
\tikzstyle{diamond marker}=[mark=diamond*, mark options={solid, fill=white, mark size=2.0pt}]
\tikzstyle{triangle marker}=[mark=triangle*, mark options={solid, fill=white, mark size=2.0pt}]
\tikzstyle{square marker}=[mark=square*, mark options={solid, fill=white, mark size=1.3pt}]
\tikzstyle{circle marker}=[mark=*, mark options={solid, fill=white, mark size=1.5pt}]
\tikzset{%
	partial ellipse/.style args={#1:#2:#3}{%
		insert path={+ (#1:#3) arc (#1:#2:#3)}%
	}%
}%

\usepackage{cite}

\newtheorem{remark}{Remark}

\newcommand{\vect}[1]{\ensuremath{\boldsymbol{#1}}}
\newcommand{\mat}[1]{\ensuremath{\bm{#1}}}
\newcommand{\imag}{\jmath}

\newcommand{\Ftot}{F_{\text{tot}}}

\makeatletter
\DeclareRobustCommand{\rvdots}{%
  \vbox{
    \baselineskip4\p@\lineskiplimit\z@
    \kern-\p@
    \hbox{.}\hbox{.}\hbox{.}
  }}
\makeatother

\newsavebox{\eyeMat}
\savebox{\eyeMat}{$\left[\begin{smallmatrix}1&0\\0&1\end{smallmatrix}\right]$}

\DeclareFontFamily{OT1}{pzc}{}
\DeclareFontShape{OT1}{pzc}{m}{it}{<-> s * [1.10] pzcmi7t}{}
\DeclareMathAlphabet{\mathpzc}{OT1}{pzc}{m}{it}

\pgfplotsset{compat=newest}%

\definecolor{DarkGreen}{rgb}{0.0, 0.5, 0.0}

\newcommand{\RemovedMath}[1]{%
    \ifmmode\text{\sout{\ensuremath{#1}}}\else\sout{#1}\fi%
}%

\newcommand{\NoRev}[1]{%
	{
	#1%
	}%
}%

\newcommand{\RevA}[1]{%
	{
	#1%
	}%
}%

\newcommand{\RevB}[1]{%
	{
	#1%
	}%
}%

\newcommand{\isthesis}{0}

\begin{document}
\bstctlcite{IEEEexample:BSTcontrol}

\title{Model-Based Machine Learning for Joint Digital Backpropagation and PMD Compensation}

\ifthenelse{\isthesis=1}{%
    \author{Rick M.~B\"utler}%
    \date{\today}%
    \IEEEspecialpapernotice{(MSc Thesis)}%
}{%
    \author{
      Rick M.~B\"utler,
      Christian H\"ager,
      Henry D.~Pfister, 
      Gabriele Liga,
      and Alex Alvarado
    \thanks{%
        This work is part of a project that has received funding from the European Union's Horizon 2020 research and innovation programme under the Marie Sk\l{}odowska-Curie grant agreement No.~749798. The work of H.~D.~Pfister was supported in part by the National Science Foundation (NSF) under Grant No.~1609327. The work of A.~Alvarado and G.~Liga has received funding from the European Research Council (ERC) under the European Union's Horizon 2020 research and innovation programme (grant agreement No.~757791). This work is also supported by the NWO via the VIDI Grant ICONIC (project number 15685). Any opinions, findings, recommendations, and conclusions expressed in this material are those of the authors and do not necessarily reflect the views of these sponsors.
    
        R.~M.~B\"utler, G.~Liga, and A.~Alvarado are with the Information and Communication Theory Lab, Signal Processing Systems Group, Department of Electrical Engineering, Eindhoven University of Technology, Eindhoven, The Netherlands (e-mails: rick-butler@outlook.com, \{g.liga, a.alvarado\}@tue.nl).
        
        C.~H\"ager is with the Department of Electrical Engineering, Chalmers University of Technology, Gothenburg, Sweden (e-mail: christian.haeger@chalmers.se).
        
        H.~D.~Pfister is with the Department of Electrical and Computer
    	Engineering, Duke University, Durham, USA (e-mail: henry.pfister@duke.edu).
    }%
    }%
}

\maketitle

\begin{abstract}
In this paper, we propose a model-based machine-learning approach for dual-polarization systems by parameterizing the split-step Fourier method for the Manakov-PMD equation. The resulting method combines hardware-friendly time-domain nonlinearity mitigation via the recently proposed learned digital backpropagation (LDBP) with distributed compensation of polarization-mode dispersion (PMD). We refer to the resulting approach as LDBP-PMD. We train LDBP-PMD on multiple PMD realizations and show that it converges within $1\%$ of its peak dB performance after $428$ training iterations on average, yielding a peak effective signal-to-noise ratio of only $0.30$~dB below the PMD-free case. Similar to state-of-the-art lumped PMD compensation algorithms in practical systems, our approach does not assume any knowledge about the particular PMD realization along the link, nor any knowledge about the total accumulated PMD. This is a significant improvement compared to prior work on distributed PMD compensation, where knowledge about the accumulated PMD is typically assumed. We also compare different parameterization choices in terms of performance, complexity, and convergence behavior. Lastly, we demonstrate that the learned models can be successfully retrained after an abrupt change of the PMD realization along the fiber. 

\end{abstract}


\IEEEpeerreviewmaketitle

\section{Introduction}
\label{sec:introduction}

The rapidly increasing demand for reliable long-distance data transmissions presents a need for increased capacity in optical communication links \cite{Monroe2016}. One of the main limiting factors for capacity in optical channels is the nonlinear Kerr effect imposed on signals during propagation \cite{Stolen1973}. The nonlinearity mixes with linear effects like chromatic dispersion (CD) \cite{Kao1966,Nagel1989} and polarization-mode dispersion (PMD) \cite{Ip2010a,Menyuk2004b}, which makes it challenging to develop suitable schemes that can invert the nonlinear transfer function from channel input to output. In this paper, we consider the application of deep learning as a potential solution for designing powerful nonlinear signal compensation algorithms, which has recently gained significant attention in the literature \cite{Shen2011, Jarajreh2015, Estaran2016, OShea2017, Shen2018ecoc, Jones2018, Karanov2018, Khan2019}.

The traditional application of deep learning to physical-layer communication replaces or augments digital signal processing (DSP) blocks (e.g., equalization or decoding) with artificial neural networks (ANNs). An ANN consists of many concatenated linear and nonlinear operators that can be optimized to approximate a desired input--output function. The aim of this approach is to learn new algorithms through data-driven optimization that can perform better than the conventional DSP blocks, e.g., in terms of accuracy or computational load. More generally, one may regard the entire transceiver design as an end-to-end reconstruction task and jointly optimize transmitter and receiver ANNs \cite{OShea2017}. Both traditional \cite{Shen2011, Jarajreh2015, Estaran2016} and end-to-end learning \cite{Shen2018ecoc, Jones2018, Karanov2018} have received considerable attention for optical fiber systems. However, ANNs are often used as 
``black boxes'' which makes it difficult to incorporate existing domain knowledge. As opposed to conventional DSP algorithms which are based on well-understood mathematical models and theory, the structure of the algorithm learned by an ANN has no obvious relation to the problem at hand and their high-level operation is difficult to interpret.

Rather than relying on ANNs, a different approach is to start from an existing model and parameterize it. For fiber-optic systems, it has been shown in \cite{Haeger2018ofc} that this can be done by considering the split-step Fourier method (SSFM) for numerically solving the nonlinear Schr\"{o}dinger equation (NLSE). By viewing all CD steps as general linear functions, one obtaines a parameterized model similar to a multi-layer ANN \cite{Haeger2018ofc}. This approach has several advantages over general-purpose ANNs \cite{Haeger2018ofc, Haeger2018isit, Lian2018itw, Haeger2019ecoc, Oliari2020}: it allows for a more intuitive selection of hyperparameters such as the number of layers; it provides good initializations for gradient-based optimization; and it allows one to inspect and interpret the obtained solutions in order to understand \emph{why} they work well, thereby providing significant insight into the problem.

The approach in \cite{Haeger2018ofc} only applies to the standard NLSE and does not take polarization-dependent effects into account. This paper is an expanded version of \cite{Haeger2020ofc}, where our main contribution is to extend the model-based machine-learning approach proposed in \cite{Haeger2018ofc} to dual-polarization (DP) transmission by parameterizing the SSFM for the Manakov-PMD equation. The proposed parameterization leads to a multi-layer model alternating complex-valued $2 \times 2$ multiple-input multiple-output finite impulse reponse (MIMO-FIR) filters with nonlinear Kerr operators. The complexity of the MIMO filters is reduced by decomposing them into separate FIR filters for each polarization followed by memoryless rotation matrices \cite{Haeger2019ecoc, Oliari2020}. This decomposition mimics the forward propagation model, where PMD introduces a polarization-dependent differential group delay (DGD) and rotates the principal states of polarization (PSP) along the fiber link in a distributed fashion \cite{Nelson2005}.

As an application of the proposed machine-learning model, we consider joint digital backpropagation (DBP) and distributed PMD compensation. This is similar to \cite{Goroshko2016, Czegledi2016a, Czegledi2017, Liga2018}, where PMD-compensating sections are inserted into the standard DBP algorithm. However, all of these works assume that the accumulated PMD is known to the receiver, either to heuristically compute the individual PMD sections \cite{Goroshko2016, Czegledi2016a, Czegledi2017} or to initialize a gradient-free optimization procedure \cite{Liga2018}. On the other hand, our approach relies on gradient-based deep learning and does not assume any knowledge about the particular PMD realizations along the link (i.e., the DGDs and PSPs), nor any knowledge about the total accumulated PMD. Moreover, the employed model uses hardware-friendly time-domain implementations \cite{Fougstedt2017, Fougstedt2018ecoc} based on very short FIR filters to account for DGD. We demonstrate that our model converges reliably to a solution whose performance is close to the case where PMD is absent from the link. 

Compared to the conference version \cite{Haeger2020ofc}, this paper aims to provide a thorough numerical investigation of different parameterization choices including different initialization schemes, by comparing them in terms of performance, complexity, and convergence behavior. We also show that our method outperforms the case where a single lumped MIMO-FIR filter is learned. Finally, we demonstrate that the learned model can be successfully retrained after an abrupt change of the PMD realization along the fiber, which constitutes an important first step towards an adaptive real-time implementation of the proposed approach.

\section{Deep Learning Review} \label{sec:deepLearning}

In this section, we give a brief overview of the general theory behind deep learning.

\subsection{Artificial Neural Networks}
\label{sec:deepLearning:structure}

An ANN is a directed computational graph where each node $n$ applies a scalar function $f^{(n)}(\vect{x}^{(n)};\vect{\Psi}^{(n)})$ to its input vector $\vect{x}^{(n)}$ \cite[Ch.~6]{Goodfellow2016}, \cite[Ch.~1]{Nielsen2015}. Here, $\vect{\Psi}^{(n)}$ is a set of parameters that, in combination with the analytical definition of $f^{(n)}$, defines the node operation. An ANN node usually applies a linear operation to propagate each of its inputs to its output, followed by a nonlinear activation function. An example of an often-used node operation is
\begin{equation} \label{eq:MLP}
    f^{(n)}(\vect{x}^{(n)};\vect{\Psi}^{(n)})=\varepsilon((\vect{w}^{(n)})^\top\vect{x}^{(n)}+b^{(n)}),
\end{equation}
where $\varepsilon$ is the activation function, $\vect{w}^{(n)}$ is a weight vector, $b^{(n)}$ is a bias and $\vect{\Psi}^{(n)}=\{\vect{w}^{(n)},b^{(n)}\}$. 

\begin{figure}[t]
    \centering%
    \includegraphics[width=0.9\columnwidth]{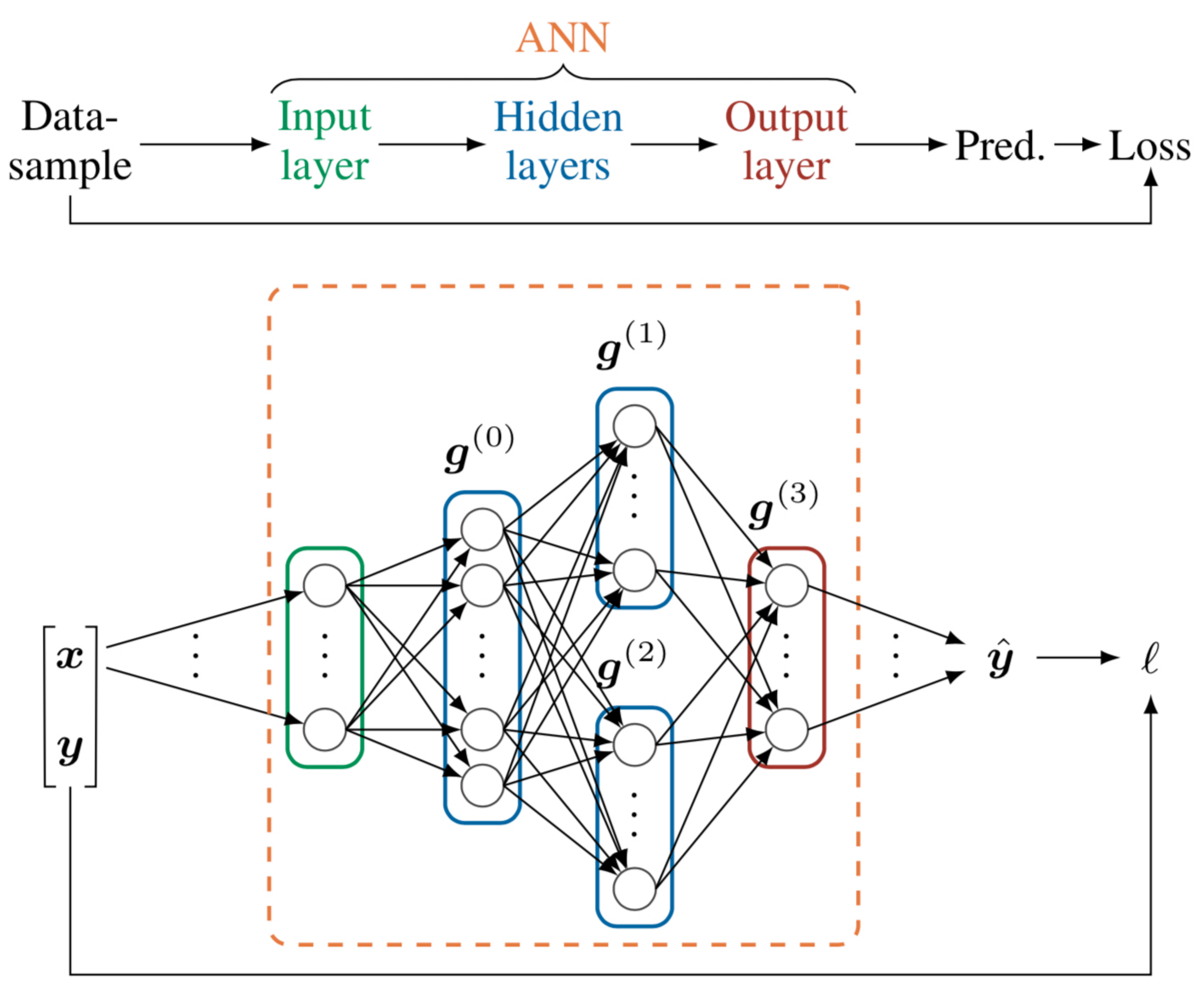}
    \caption{Example feed-forward ANN with fully-connected  layers.}%
    \vspace{-0.25cm}%
    \label{fig:ANN}%
\end{figure}%

Groups of nodes that share the same inputs are abstracted into MIMO functions
\begin{equation} \label{eq:layer}
    \vect{g}^{(l)}(\vect{x}^{(l)};\vect{\Phi}^{(l)})=[f^{(0)}(\vect{x}^{(l)};\vect{\Psi}^{(0)}),\cdots, f^{(N\mbox{-}1)}(\vect{x}^{(l)};\vect{\Psi}^{(N\mbox{-}1)})],
\end{equation}
where $\vect{\Phi}^{(l)}=\{\vect{\Psi}^{(0)},\cdots,\vect{\Psi}^{(N\mbox{-}1)}\}$ and nodes $0$ through $N\mbox{-}1$ form layer $l$. An ANN is constructed by defining $L$ layers and connecting their in- and outputs to obtain a large overall directed computational graph $\vect{y} = f(\vect{x};\vect{\theta}) \in \mathcal{Y}$, where $\vect{x} \in \mathcal{X}$ is the input, $\mathcal{X}$ and $\mathcal{Y}$ are the input and output alphabets (typically sets of vectors), respectively, and  $\vect{\theta}=\{\vect{\Phi}^{(0)},\cdots,\vect{\Phi}^{(L-1)}\}$ collects all tunable parameters. An example of a feed-forward ANN with fully-connected layers is shown in Fig.~\ref{fig:ANN}.

\subsection{Gradient-Based Learning}  \label{sec:deepLearning:optimization}

The parameter set $\vect{\theta}$ can be optimized in an iterative fashion with data-driven gradient-based optimization methods as follows. Given a data set $D \subset \mathcal{X} \times \mathcal{Y}$, a minibatch $B_\kappa \subseteq D$ is sampled from $D$, where $|B_\kappa|$ is the minibatch size, and $\kappa$ is the iteration index. For each input--output pair $(\vect{x}, \vect{y}) \in B_\kappa$, the input $\vect{x}$ is propagated through the ANN to obtain a prediction $\hat{\vect{y}}=f\left(\vect{x};\vect{\theta}\right)$. The prediction is then compared to the desired output $\vect{y}$ by using a per-sample loss function $\ell(\hat{\vect{y}}, \vect{y})$. The average loss associated with the entire minibatch is given by 
\begin{equation}
    \mathcal{L}_{B_\kappa}(\vect{\theta}) = \frac{1}{|B_\kappa|}\sum_{(\vect{x}, \vect{y}) \in B_\kappa }\ell(f\left(\vect{x};\vect{\theta}\right), \vect{y}).
\end{equation}%
Afterwards, the gradient $\nabla_{\vect{\theta}}\mathcal{L}_{B_\kappa}(\vect{\theta})$ with respect to the network parameters is calculated numerically, by applying the chain rule to local gradients per node \cite[Ch.~2]{Nielsen2015}. Finally, the computed gradient is used to take an optimization step with a chosen optimizer. 

The simplest optimizer is stochastic gradient descent (SGD), which applies 
\begin{align}
\vect{\theta}_{\kappa+1}=\vect{\theta}_\kappa-\eta\nabla_{\vect{\theta}}\mathcal{L}_{B_\kappa}(\vect{\theta}),
\end{align}
where $\eta$ is a chosen learning rate. Due to problems like small gradients at subobtimal values of $\vect{\theta}$ (e.g., saddle points or flat regions), and variance in the sensitivity of $\mathcal{L}$ to different parameters, SGD is often not enough to effectively navigate the loss landscape. In this work, we rely on the Adam optimizer \cite{Kingma2014a} that addresses both mentioned issues by introducing momentum and per-parameter adaptive learning rates.

The combination of linearities and nonlinearities in ANNs allows for the learning of complex functions. On the other hand, a downside of deep learning with ANNs is that, although ANNs can potentially converge to very accurate solutions, these solutions are generally hard to interpret. In this paper, this issue is countered by implementing a model whose structure is based on a modified version of the SSFM, which is a well-understood existing algorithm. Indeed, supervised learning is not restricted to ANNs and learning algorithms such as SGD can be applied to other function classes as well. For example, in state-of-the-art practical systems, PMD is typically compensated by choosing the function $f(\cdot, \vect{\theta})$ as the convolution of the received signal with the impulse response of a complex-valued $2 \times 2$ MIMO-FIR filter, where $\bm{\theta}$ are the filter coefficients. For a particular choice of the loss function $\ell$, applying SGD then maps into the well-known constant modulus algorithm \cite{Savory2008}.

\section{Model-Based Machine Learning for Dual-Polarization
Systems} \label{sec:modelBasedLearning}

\subsection{The Manakov-PMD Equation} \label{sec:modelBasedLearning:manakov}
The evolution of DP signals through a single-mode fiber can be described by a set of coupled NLSEs that takes into account the interactions between the two degenerate polarization modes as a function of propagation distance \cite{Nelson2005}. In birefringent fibers where the PSP changes rapidly along the link, propagation in a noiseless fiber is governed by the Manakov-PMD equation \cite{Wai1991}
\begin{equation} \label{eq:manakovpmd}
\begin{aligned}
    \frac{\partial\vect{u}}{\partial z} &= \left(-\frac{1}{2}\alpha-\mat{\beta}_1(z)\frac{\partial}{\partial t}-\jmath\beta_2\frac{1}{2}\frac{\partial^2}{\partial t^2}\right)\vect{u}+\jmath\frac{8}{9}\gamma\|\vect{u}\|^2\vect{u},
\end{aligned}
\end{equation}
where $\vect{u} = \vect{u}(t,z) = [u_x(t,z), u_y(t,z)]^\top$ is the Jones vector comprising the complex baseband signals in both polarizations, $t$ and $z$ are propagation time and distance, $\alpha$ is fiber attenuation, $\mat{\beta}_1(z)\in\mathbb{C}^{2\times 2}$ represents PMD, $\beta_2$ is group-velocity dispersion giving rise to CD, dispersion slope $\beta_3$ is neglected, $\gamma$ is the nonlinear Kerr parameter, $\|\vect{u}\|=\sqrt{\vect{u}^\text{H}\vect{u}}$, and $\square^\text{H}$ denotes the Hermitian transpose. $\mat{\beta}_1(z)$ can be decomposed into DGD and PSP rotations, which will be discussed in detail in the following subsection.

\subsection{Distributed PMD Modeling in the SSFM} \label{sec:modelBasedLearning:PMDModeling}

\begin{figure}[tb]
\centering
\includegraphics[width=0.9\columnwidth]{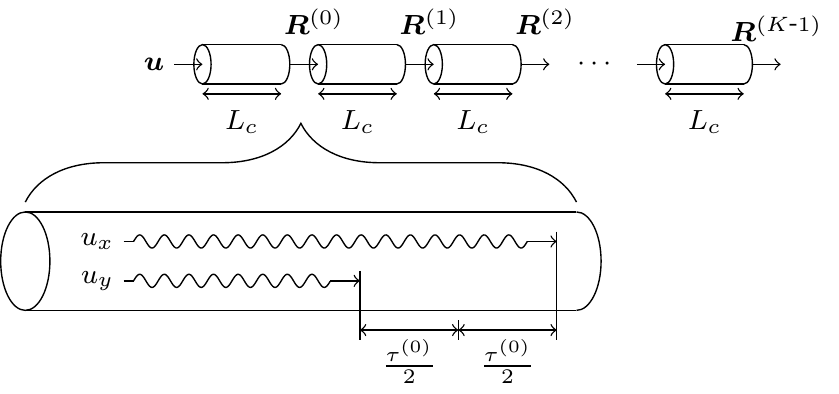}
\caption{A fiber divided into $K$ length-$L_c$ PMD sections. PSP-rotating matrices $\mat{R}^{(k)}$ are applied after each section. DGD is applied during each section, where $\tau^{(k)}$ is defined as the delay section $k$ adds between the polarizations.} \label{fig:PMDSections}.
\end{figure}

The correlation length $L_c$ of a fiber is defined as the distance which a DP signal can travel before its PSP is uncorrelated with its initial state \cite{Wai1991, Menyuk2004b}. In order to model PMD, the fiber is divided into $K$ PMD sections of length $L_c$ as illustrated in Fig.~\ref{fig:PMDSections}, where each section $k=0,\cdots,K-1$ applies a PMD operator
\begin{equation} \label{eq:beta1}
\mat{\beta}_1^{(k)}=\mat{R}^{(k)}\left(\frac{\tau^{(k)}}{2}\mat{\sigma}_1\right), 
\end{equation}%
where $\tau^{(k)}\in\mathbb{R}$ is the DGD that section $k$ introduces between the two polarizations, $\mat{\sigma}_1=\left[\begin{smallmatrix}1&0\\0&-1\end{smallmatrix}\right]$ is a Pauli spin matrix, $\mat{R}^{(k)} \in \text{SU}(2)$ is a PSP rotation matrix, and
\begin{align}
    \label{eq:su2}
    \text{SU}(2) = 
    \left\{%
    \begin{bmatrix}a&b\\-b^*&a^{*}\end{bmatrix} : a,b\in\mathbb{C}, |a|^2+|b|^2=1
    \right\},%
\end{align}
is the special unitary group of degree $2$. In the context of \eqref{eq:manakovpmd} with $\alpha = \beta_2 = \gamma = 0$, the application of the DGD operator can be solved in the frequency domain, which results in the application of a frequency-dependent Jones matrix%
\begin{equation} \label{eq:DGDMat}
    \mat{J}^{(k)}(\omega)=\begin{bmatrix}\exp\left(-\jmath\omega\frac{\tau^{(k)}}{2}\right)&0\\0&\exp\left(\jmath\omega\frac{\tau^{(k)}}{2}\right)\end{bmatrix}.
\end{equation}%
Thus, each PMD section $k$ applies a Jones matrix $\mat{R}^{(k)}\mat{J}^{(k)}(\omega)$ to its frequency-transformed input, where
\begin{align}
    \mat{J}(\omega) = \prod_{k=0}^{K-1} \mat{R}^{(k)}\mat{J}^{(k)}(\omega)
\end{align}
is the overall PMD transfer function matrix. 

Like the NLSE, the Manakov-PMD equation has no general closed-form solution, but it can be solved numerically using the SSFM \cite[Ch.~2]{Agrawal2019}, \cite[Ch.~1]{Ellis2019}. This is done by merging the step distribution of the SSFM for the NLSE in each span with $K$ uniform PMD sections of length $L_c$. During each SSFM step of length $h$ within PMD section $k$, a fraction of the DGD $\tau^{(k)}h/L_c$ is applied by substituting it for $\tau^{(k)}$ in \eqref{eq:DGDMat}. After the last SSFM step in the $k$-th PMD section, the PSP is rotated by $\mat{R}^{(k)}$. Additionally, CD is modeled like in the regular SSFM for the NLSE as
\begin{equation}
    \tilde{\vect{u}}\exp\left(\jmath\frac{1}{2}\beta_2\omega^2h\right),
\end{equation}
where $\tilde{\boldsymbol{u}}=\tilde{\boldsymbol{u}}(\omega,z)$ is the baseband signal in the frequency domain, and the Kerr effect is modeled as
\begin{equation}
    \vect{u}\exp\left(\jmath\frac{8}{9}\gamma h\|\vect{u}\|^2\right).
\end{equation}

For generating a particular PMD realization, i.e., particular values for the DGDs $\tau^{(k)}$ and rotation matrices $\mat{R}^{(k)}$ for all sections $k = 0, \cdots, K-1$, we use the same approach as described in \cite{Czegledi2016a}. More specifically, $\tau^{(k)}$ is sampled from a Gaussian distribution with mean $\overline{\tau}$ and standard deviation $\overline{\tau}/5$, where $\overline{\tau}=\tau\sqrt{3\pi/8\cdot L_c}$ and $\tau$ is the PMD parameter. PSP rotation matrices are generated randomly and uniformly from the set SU(2) such that the PSP after being rotated is uniformly distributed over the surface of the Poincaré sphere, see, e.g., \RevB{\cite[p.~7]{Czegledi2017}}.

\subsection{Proposed Machine-Learning Model} \label{sec:modelBasedLearning:PMDCompensation}

\begin{figure}[t]
	\centering
    \includegraphics[width=0.8\columnwidth]{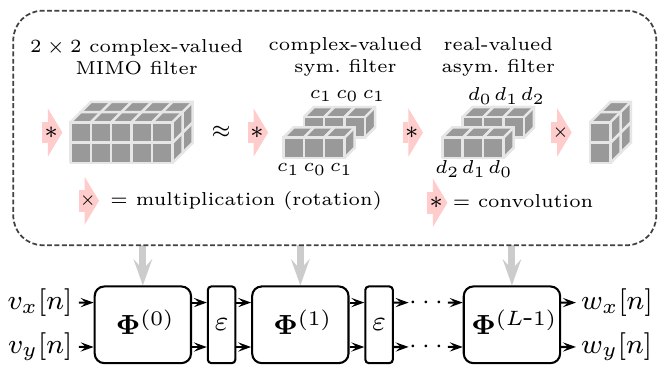}
	\caption{Block diagram of the proposed model \RevA{$\vect{w}[n]=f(\vect{v}[n];\vect{\theta})$ based on the SSFM for the Manakov-PMD equation, where $\vect{v}[n] = [v_x[n], v_y[n]]^\top$ is the sequence of receiver inputs in both polarizations after downsampling, $\vect{w}[n] = [w_x[n], w_y[n]]^\top$ is the model output, and} all filter coefficients are free parameters $\vect{\theta}= \{\mat{\Phi}^{(0)},\cdots,\mat{\Phi}^{(L-1)}\}$. 
	The framed box illustrates the proposed MIMO filter decomposition, where the individual filters have length $3$ as an example. }
	\label{fig:model}
\end{figure}

For DP signals, each linear step in the SSFM corresponds to the combined effect of all linear operators in the Manakov-PMD equation \eqref{eq:manakovpmd}, which can be approximated by a $2\times 2$ complex MIMO-FIR filter in discrete time. The proposed approach is to fully parameterize these MIMO filters by regarding all filter coefficients as free parameters. Similar to an ANN, this gives a parameterized multi-layer model alternating linear steps and nonlinearities according to
\begin{equation}
    \label{eq:Kerr}
    {\varepsilon}(\vect{u})=\vect{u}\exp\left(\jmath\frac{8}{9}\gamma h\|\vect{u}\|^2\right),
\end{equation}
where a corresponding block diagram is shown in the bottom of Fig.~\ref{fig:model}.
\RevA{The model can be written as $\vect{w}[n] = f(\vect{v}[n];\vect{\theta})$, where $\vect{v}[n] = [v_x[n], v_y[n]]^\top$ is the sequence of receiver inputs in both polarizations after downsampling and $\vect{w}[n] = [w_x[n], w_y[n]]^\top$ is the model output.}
In this paper, we focus on receiver-side DBP as an application, where the model can learn to apply arbitrary filter shapes to DP signals in a distributed fashion. This essentially extends the learned DBP (LDBP) approach proposed in \cite{Haeger2018ofc} to DP signals and we refer to the resulting method as LDBP-PMD. In general, the model can also be used at the transmitter to learn optimized pre-distortions \cite{Essiambre2005, Roberts2006} or split-DBP \cite{Lavery2016a}.

For multi-layer models, it is important to simplify the individual steps as much as possible to limit the overall complexity \cite{Haeger2019ecoc, Oliari2020}. This is especially important since PMD is a time-varying impairment that requires adaptive filtering in practice \cite{Nelson2005}. We therefore propose to simplify the individual linear steps by decomposing the MIMO-FIR filter into three components as illustrated in the top part of Fig.~\ref{fig:model}:
\begin{enumerate}
\item Two complex-valued symmetric filters of odd length $F_k'$ that can account for CD in each step, where the same filter taps $\vect{c}^{(k)} = [c_{\frac{F_k'-1}{2}}^{(k)}, \cdots, c_0^{(k)}, \cdots, c_{\frac{F_k'-1}{2}}^{(k)} ]$ are used in both polarizations. 

\item Two real-valued asymmetric filters of length $F_k$ with ``flipped'' coefficients $\vect{d}^{(k)} = [d_0^{(k)},d_1^{(k)},\cdots, d_{F_k-1}^{(k)}]$ in different polarizations to approximate fractional-delay DGD filters. In this paper, all DGD filters are assumed to have the same length, denoted by $F$ in the following. 

\item A complex-valued $2\times 2$ PSP-rotating Jones matrix.

\end{enumerate}
\RevA{Note that the model output at (discrete) time $n$ depends on the model input at times $n-\frac{\Ftot-1}{2}$, ..., $n+\frac{\Ftot-1}{2}$, where $\Ftot$ is the total memory introduced by all filters in the model, i.e., 
\begin{align}
   \Ftot = \sum_{k=0}^{K-1} (F_k' + F_k) - 2K,
\end{align}
and the term $-2K$ accounts for the total length when convolving $2K$ filters.}
We also note that attenuation is implicitly taken into account through a scaling of the filter coefficients in the linear steps. Alternatively, one may introduce a scaling factor in the computation of the nonlinear phase shift in \eqref{eq:Kerr}.

\subsection{Parameterizations}
\label{sc:parameterizations}

We consider multiple parameterizations for the compensating DGD filters and PSP rotating matrices, which result in varying degrees of freedom (DOF) for the optimization and computational load. For the DGD filters, we consider the following two cases:
\begin{enumerate}
    \item All filter taps are freely optimized, which gives $F$ DOF, where $F$ is the filter length. 
    
    \item The DGD filters are constrained as Lagrange fractional-delay filters with filter taps according to \cite{Kootsookos1996}%
    \begin{equation}
    \label{eq:lagrange}
    d_n^{(k)}=\prod_{i=0,i\neq n}^{F\mbox{-}1}\frac{f_s\tau^{(k)}/2-i}{n-i},
    \end{equation}%
    where $f_s$ is the sampling frequency. In this case, the DGD $\tau^{(k)}$ is optimized directly, resulting in only $1$ DOF per model step, independently of the filter length $F$. 
    
\end{enumerate}
For both of the above parameterizations, applying the two real-valued filters results in $4F$ real multiplications (RMs) per $2$ polarizations and model step. 

For the PSP rotation matrices, we also consider two parameterizations, both of which require $16$ RMs per model step:
\begin{enumerate}
    \item All $4$ complex-valued matrix entries are freely optimized, which results in $8$ DOF. 
    
    \item The matrices are from SU(2), but without imposing a unitary constraint on the rotation. We denote the resulting set as
    \begin{align}
        \text{SU}(2)^* = 
        \left\{%
        \begin{bmatrix}a&b\\-b^*&a^{*}\end{bmatrix} : a,b\in\mathbb{C}
        \right\},%
    \end{align}
    which reduces the number of DOF to $4$. 
    
\end{enumerate}
The complexity of each parameterization in terms of DOF and RMs per model step is shown in Table~\ref{tab:dof}. The case of a full $2 \times 2$ MIMO-FIR filter of length $F$ is also shown as reference. 

\begin{table}[t]
    \centering
    \caption{Degrees of Freedom (DOF) and real multiplications (RMs) per $2$ polarizations for each model step with filter length $F$.}
    \label{tab:dof}
    \begin{tabular}{c|c|c}
        \textbf{Parameterization}&\textbf{DOF}&\textbf{RMs}\\
        \hline\hline
        free MIMO filter&$8F$&$16F$\\
        &&\\
        \textbf{DGD}&&\\
        \hline
        free DGD filter&$F$&$4F$\\
        Lagrange filter&$1$&\\
        &&\\
        \textbf{PSP rotation}&&\\
        \hline
        free matrix&$8$&$16$\\
        SU(2)$^*$ matrix&$4$&
    \end{tabular}
\end{table}

\begin{remark}
Another possible parameterization for the rotation matrices that is used, for example, in \cite{Liga2018} is via an alternative definition of SU(2) according to
\begin{equation} \label{eq:RPhi}
\text{SU}(2) = \left\{e^{-\jmath({\phi_1\mat{\sigma}_1+\phi_2\mat{\sigma}_2+\phi_3\mat{\sigma}_3})} : \phi_1, \phi_2, \phi_3 \in \mathbb{R} \right\},
\end{equation}%
where $\mat{\sigma}_2=\left[\begin{smallmatrix}0&1\\1&0\end{smallmatrix}\right]$ and $\mat{\sigma}_3=\left[\begin{smallmatrix}0&-\jmath\\\jmath&0\end{smallmatrix}\right]$ are Pauli spin matrices, and $\phi_1,\phi_2,\phi_3$ are rotation angles in Stokes space. While this further reduces the DOF to $3$, gradient-based optimization based on \eqref{eq:RPhi} is challenging due to the matrix exponential which is currently not supported in the deep-learning framework used for our implementation (PyTorch). 
\end{remark}

\section{Simulation Setup} \label{sec:setup}

\begin{table}[tb]
\centering
\caption{Transceiver \& channel properties.} \label{tab:constants}
\begin{tabular}{c|c}
\textbf{Property}&\textbf{Value}\\
\hline\hline
constellation&Gaussian\\
pulse shape&1\% root-raised cosine\\ 
symbol rate&$32~\text{Gbaud}$\\
forward SSFM bandwidth&$192~\text{GHz}$ ($6$ samples/symbol)\\
receiver bandwidth&$64~\text{GHz}$ ($2$ samples/symbol)\\
fiber spans&$10\times100~\text{km}$\\
SSFM StPS&$1000$ linear asymmetric\\
DBP StPS&$1000$ mod-logarithmic symmetric \cite{Zhang2013}\\
LDBP/LDBP-PMD StPS&$4$ mod-logarithmic symmetric\\
$\alpha$&$0.2~\text{dB}/\text{km}$\\
$\beta_2$&$-21.67~\text{ps}/\text{km}^2$\\
$\gamma$&$1.2~\text{rad}/\text{W}/\text{km}$\\
$\tau$&$0.2~\text{ps}/\sqrt{\text{km}}$\\
$L_c$&$0.1~\text{km}$\\
amplifier noise figure&$4.5~\text{dB}$\\
central wavelength&$1550~\text{nm}$
\end{tabular}
\end{table}

\begin{figure}
    \centering
	 \includegraphics[width=5cm]{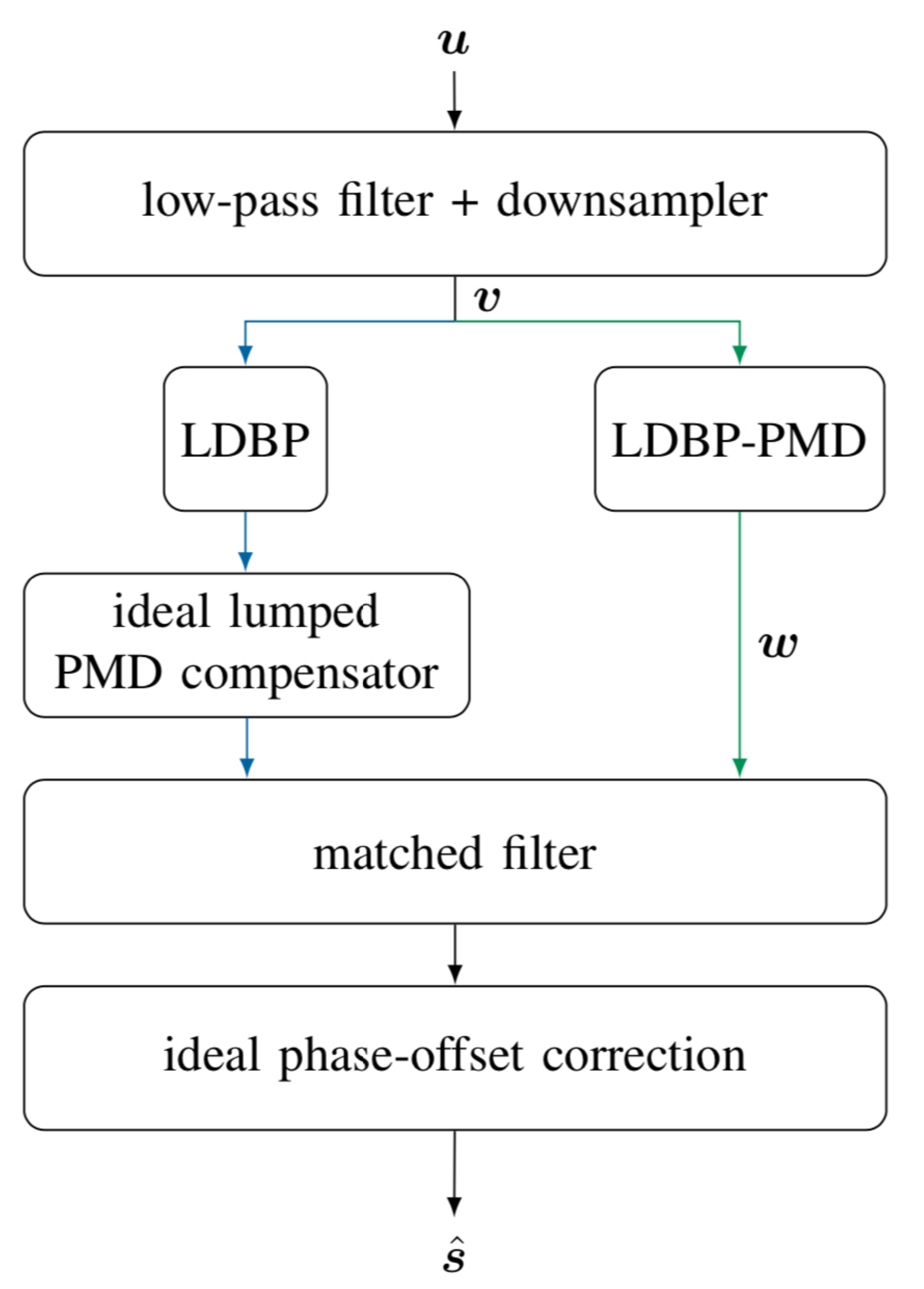}
    \caption{Block diagram of the used receiver, where $\vect{u}$ is the channel output and $\hat{\vect{s}}$ is an estimate of the transmitted symbols. The colored paths show two different compensation schemes with LDBP (blue) or LDBP-PMD (green) in the presence of PMD in the channel.}
    \label{fig:Rx}
\end{figure}

We consider the simulation setup in \cite{Czegledi2016a} with slightly adjusted parameters which can be viewed in Table~\ref{tab:constants}. For the forward simulation, each span is divided into $1000$ uniform steps per span (StPS) and an optical amplifier is inserted after each span. Note that in this case, the SSFM step size is the same as the PMD correlation length. PMD is emulated at every SSFM step as described in Sec.~\ref{sec:modelBasedLearning:PMDModeling}, where $40$ different PMD realizations, i.e., sets of $\tau^{(k)}$ and $\mat{R}^{(k)}~\forall~k\in\{0,\cdots,K-1\}$ are used to obtain the numerical results.

A block diagram of the used receiver is shown in Fig.~\ref{fig:Rx}. The signal is first low-pass filtered and downsampled. Then, either LDBP or LDBP-PMD is applied as described in the following two subsections. \RevA{We note that for the simulations, finite-length sequences are used and all linear filtering steps are implemented as circular convolutions. In this case, the input (and similarly the output) of the machine-learning models are two vectors of length $2 N_\text{sym}$ (one vector for each polarization), where $N_\text{sym}$ is the number of symbols per polarization and $2$ is the oversampling factor used at the receiver.} Finally, a matched filter and ideal phase-offset compensation are applied. The phase-offset correction is genie-aided according to
$\hat{\vect{s}} = \tilde{\vect{s}} e^{-\imag \hat{\phi}}$, where
$\hat{\phi} = \arg ( \vect{s}^\text{H} \tilde{\vect{s}}) $,
$\tilde{\vect{s}}$ is the symbol vector after the matched filter, and $\vect{s}$ is the originally transmitted symbol sequence. The accuracy of the receiver output sequence $\hat{\vect{s}}$ is evaluated in terms of the effective SNR defined as
\begin{align}
    \label{eq:eSNR}
    \text{effective SNR}= \frac{\|\hat{\vect{s}}\|^2}{\|\hat{\vect{s}}-\vect{s}\|^2}.
\end{align}

\subsection{LDBP Baseline}

Before optimizing LDBP-PMD, we start by training a standard $4$-StPS LDBP model, where only the complex-valued symmetric filters $\vect{c}^{(k)}$ are being optimized \cite{Haeger2018ofc}. The resulting model has a total of $41$ steps, where the last step is a purely linear ``half-step''. \RevB{Due to the use of mod-logarithmic step sizes, }the individual filter lengths $F'_k$ are different in each step, where the average filter length is $25$ (symmetric) taps. PMD is assumed absent from the link during training, i.e, $\tau^{(k)}=0$,~$\mat{R}^{(k)}=\mat{I}_2~\forall~k$. The performance of the trained standard LDBP model with and without PMD on the link is used as a reference for all further trained receivers. With PMD, a MIMO filter with frequency response $\mat{J}^{-1}(\omega)$ is applied after LDBP that perfectly inverts the PMD accumulated over the link.

\subsection{LDBP-PMD}

The trained LDBP model is then used as the starting point for all LDBP-PMD optimizations. Five different parameterizations are considered, where the first parameterization uses a MIMO-FIR filter as a reference. The other four parameterizations use the different combinations of DGD filters and PSP rotation matrices described in Sec.~\ref{sc:parameterizations}. The MIMO and DGD filter length is set to $F = 5$ in all cases. In the following, the five parameterizations will be referred to as
\begin{enumerate}
\item ``free MIMO filters'',
\item ``free DGD filters + free matrices'',
\item ``free DGD filters + SU(2)$^*$ matrices'',
\item ``Lagrange filters + free matrices'', and
\item ``Lagrange filters + SU(2)$^*$ matrices''.
\end{enumerate}
In terms of parameter initialization, the DGD filter taps are always initialized to $[0,0,1,0,0]$. For the rotation matrices, we consider two different initialization schemes, where
\begin{enumerate}
\item $\mat{R}^{(k)}=\mat{I}_2\triangleq\left[\begin{smallmatrix}1&0\\0&1\end{smallmatrix}\right]~\forall~k\in\{0,\cdots,K-1\}$, or
\item $\mat{R}^{(k)}$ is chosen uniformly at random from SU(2).
\end{enumerate}
Each parameterization is trained for both initialization methods, giving $10$ cases in total. 
All models are trained with the Adam optimizer at $P=8$ dBm transmit power using $\text{normalized MSE}=(\text{effective SNR})^{-1}$ as a loss function, $1500$ iterations, minibatch size $|\mathcal{B}_\kappa| = 50$, sequences of \RevA{$N_\text{sym}= 512$} symbols $\boldsymbol{s}$, and the learning rates in Table~\ref{tab:learningRates} which were obtained using a grid search. 
\RevA{Unlike the learning rate, the minibatch size was not optimized. However, we repeated some of the experiments assuming a different minibatch size of $25$, $75$ and $100$, without notable changes in performance.}
During training, the pre-trained complex-valued symmetric filter taps $\vect{c}^{(k)}$ of the original LDBP model (see Fig.~\ref{fig:model}) are frozen in all steps since they are assumed to account for static impairments such as CD. 

\begin{table}[tb]
    \centering
    \caption{Learning rates for the different parameterizations.}
    \begin{tabular}{c||c|c}
        &free matrices&SU(2)$^*$ matrices\\
        \hline\hline
        free DGD filters&$1e\mbox{-}3$&$5e\mbox{-}4$\\
        Lagrange filters&$5e\mbox{-}4$&$2e\mbox{-}3$\\
        \hline\hline
        free MIMO filters&\multicolumn{2}{c}{$1e\mbox{-}3$}
    \end{tabular}
    \label{tab:learningRates}
\end{table}

\section{Numerical Results} \label{sec:results}

\subsection{LDBP-PMD Initialization} \label{sec:results:initialization}

\begin{figure}[tb]
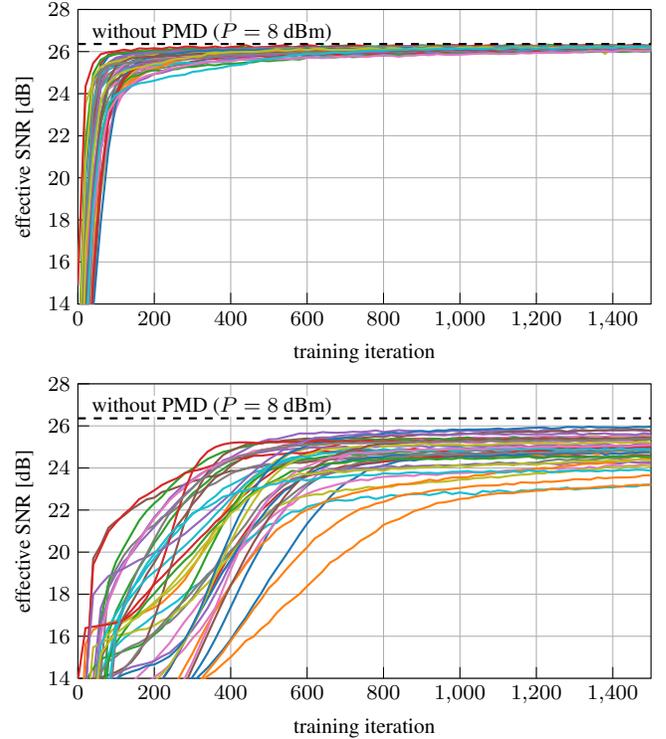
%
    \centering%
    \includestandalone{butler5top}
    \includestandalone{butler5bottom}%
    \vspace{-0.25cm}%
    \caption{Learning curves for all $40$ PMD realizations for the best (top, ``free MIMO filters'', random initialization) and worst (bottom, ``Lagrange filters + free matrices'', $\mat{I}_2$ initialization) parameterization in terms of the final mean effective SNR.}%
    \label{fig:LDBP-PMD_best_worst}%
\end{figure}

\begin{figure}[tb]
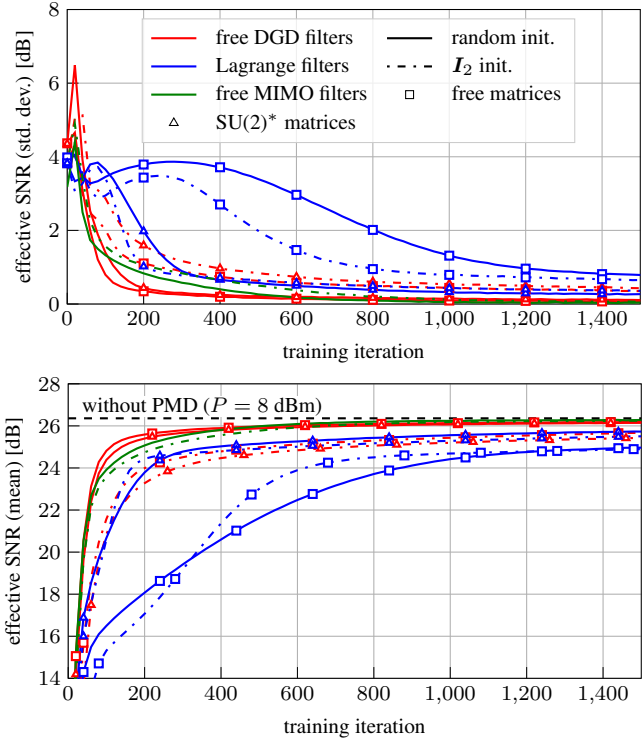

    \centering
    \includestandalone{butler6top}
    \includestandalone{butler6bottom}
    \vspace{-0.25cm}
    \caption{Effective SNR vs. training iteration for all LDBP-PMD parameterizations, trained for 40 PMD realization: standard deviation (top) and mean (bottom). All filters have length $F=5$.} \label{fig:LDBP-PMD_All_Settings_Train}
\end{figure}

\begin{figure}[tb]
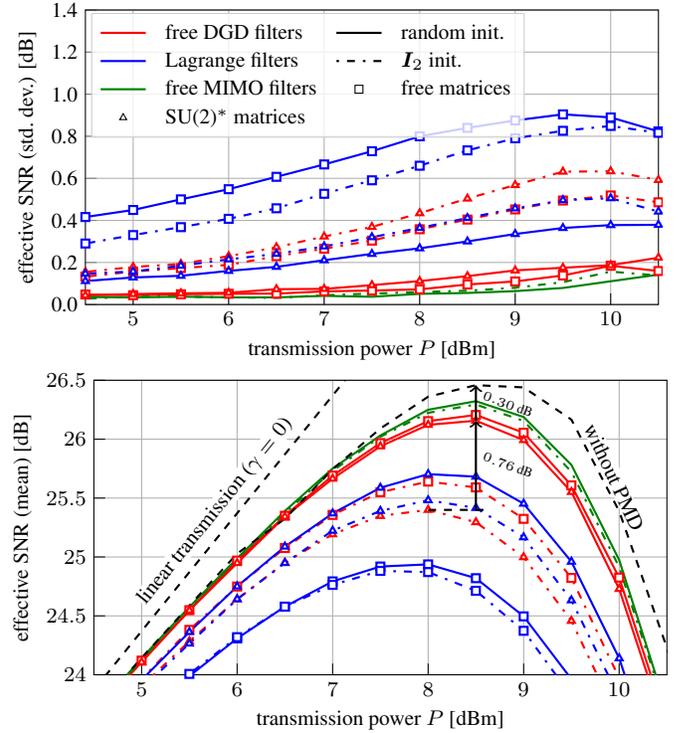
%
    \centering%
    \includestandalone{butler7top}
    \includestandalone{butler7bottom}%
    \vspace{-0.25cm}%
    \caption{Effective SNR vs. transmission power $P$ for all LDBP-PMD parameterizations, trained for 40 PMD realization: standard deviation (top) and mean (bottom). All filters have length $5$.}%
    \label{fig:LDBP-PMD_All_Settings_P}%
\end{figure}%

\RevA{In order to illustrate that the model parameterization and initialization can have a large effect on the training behavior, Fig.~\ref{fig:LDBP-PMD_best_worst} shows the $40$ individual learning curves for two combinations of parameterization and initialization methods: ``free MIMO filters'' with random initialization (top) and ``Lagrange filters + free matrices'' with $\mat{I}_2$ initialization (bottom).
It can be seen that for the first case, all learning curves exhibit a similar behavior and quickly converge to their final effective SNR value, regardless of the particular PMD realization. 
On the other hand, for the second case, the individual learning curves show a markedly different behavior with a relatively large spread in terms of the final achieved effective SNR.
The training behavior for all proposed parameterization and initialization methods is summarized in Fig.~6, where we show the mean (bottom) and standard deviation (top) of the effective SNR as a function of the training iterations. 
Note that the two cases shown in Fig.~\ref{fig:LDBP-PMD_best_worst} are those methods with the best and worst final mean effective SNR from Fig.~\ref{fig:LDBP-PMD_All_Settings_Train} after $1,500$ iterations.}
From Fig.
~\ref{fig:LDBP-PMD_All_Settings_Train}, it is clear that the way the rotations are initialized is important for the learning behavior: for all parameterizations, initializing the rotations as $\mat{I}_2$ yields a lower mean effective SNR compared to random initializations.

Fig.~\ref{fig:LDBP-PMD_All_Settings_P} shows the performance of the same cases as in Fig.~\ref{fig:LDBP-PMD_All_Settings_Train} after $1,500$ training iterations as a function of transmission power $P$, where we note that the optimization was only performed at $P=8$ dBm and the models were not retrained at other powers. \RevB{Random initializations (solid lines) outperform Identity initializations (dashed-dotted lines) in terms of mean effective SNR at optimum launch power for every parameterization. This improvement ranges from $0.028$ dB for the ``free MIMO filters'' parameterization up to $0.76$ dB for the ``free DGD filters + SU(2)$^*$ matrices'' case.} Considering that the initialization has no impact on the computational complexity, this leaves no reason to initialize the rotation matrix with $\mat{I}_2$. Therefore, all results in the remainder of this section will be obtained with random initial rotations.

\subsection{LDBP-PMD Parameterizations} \label{sec:results:parameterization}

\begin{table}[tb]
\caption{Performance metrics for all parameterizations with random initial rotations. Degrees of freedom (DOF) and real multiplications (RMs) are per step.} \label{tab:metrics}
\begin{tabular}{c||c|c|c|c|c}
    &\makecell{peak\\mean\\SNR {[dB]}}&\makecell{std.~dev.\\at peak\\SNR {[dB]}}&\makecell{mean\\conv.\\iters.}&\makecell{DOF}&\makecell{RMs}\\
    \hline\hline
    \makecell{LDBP\\w/o PMD}&$26.46$&0&-&-&-\\
    \hline
    \makecell{LDBP\\w/ PMD}&$24.28$&$0.64$&-&-&-\\
    \hline\hline
    \makecell{free\\ MIMO filters}&$26.32$&$0.055$&$443$&$40$&$80$\\
    \hline
    \makecell{free DGD filters +\\free matrices}&$26.21$&$0.095$&$361$&$13$&\multirow{4}{*}{\vspace{-1.0cm}$36$}\\
    \cline{1-5}
    \makecell{free DGD filters +\\SU(2)$^*$ matrices}&$26.16$&$0.14$&$428$&$9$&\\
    \cline{1-5}
    \makecell{Lagrange filters +\\free matrices}&$24.94$&$0.80$&$912$&$9$&\\
    \cline{1-5}
    \makecell{Lagrange filters +\\SU(2)$^*$ matrices}&$25.70$&$0.27$&$664$&$5$&\\
\end{tabular}
\end{table}

In the following, each parameterization will be assessed in terms of the following metrics:  
\begin{itemize}
    \item \emph{Performance} in terms of the peak mean effective SNR and the corresponding standard deviation,
    
    \item \emph{Complexity} in terms of the DOF and required RMs per model step, and
    
    \item \emph{Convergence speed} defined as the \NoRev{average} number of training iterations before the models converge to within $1\%$ of the final effective SNR.
    
\end{itemize}
The above metrics are summarized for each parameterization in Table~\ref{tab:metrics} assuming random initial rotations. Performance for LDBP with and without PMD in the forward link is included as a reference. The first thing to note is that ``Lagrange filters + SU(2)$^*$ matrices'' outperforms ``Lagrange filters + free matrices'' in every single metric including computational complexity. This is somewhat surprising since the free matrices offer more DOF and can hence represent a larger class of functions. A potential explanation can be obtained from Fig.~\ref{fig:LDBP-PMD_best_worst} (bottom), where it can be seen that many learning curves for the latter parameterization are still increasing, i.e., they have not yet converged. Therefore, choosing a larger number of iterations may close the gap between these two parameterizations. 
\RevA{We also note that the poor convergence speed of the Lagrange filters may be related to the fact that the functional form in \eqref{eq:lagrange} is quite restrictive.
Indeed, Lagrange filters are not guaranteed to be optimal fractional-delay filters (in a mean-squared-error sense) and they may induce larger frequency-response errors compared to the free DGD filters. 
In turn, these errors may be responsible for the slower convergence speed compared to the free DGD filters, even though fewer DOF are being optimized. }

The remaining four parameterizations present a trade-off: performance improves in terms of increased mean SNR and decreased standard deviation when computational complexity is increased. The gap in computational complexity from ``free MIMO filters'' to the next-most complex parameterization ``free DGD filters + free matrices'' is relatively large. The MIMO filters require more than twice the number of DOF and RMs while the peak mean SNR is only increased by $0.11$ dB and standard deviation decreased by a factor $1.7$. The MIMO filters might therefore be used in cases where the importance of SNR performance outweighs that of an efficient implementation or when a performance baseline is needed. The case ``Lagrange filters + SU(2)$^*$ matrices'' has the fewest number of DOF per step, which might be advantageous in a real-time implementation of the gradient-descent procedure. On the other hand, compared to the MIMO filters, the peak mean SNR is $0.62$ dB lower and the standard deviation is a factor $4.9$ higher. Both parameterizations that employ free DGD filters provide a compromise between these cases, where ``free DGD filters + free matrices'' slightly outperforms ``free DGD filters + SU(2)$^*$ matrices'' with $0.05$ dB peak mean SNR and a standard deviation decrease of factor $1.5$ at the cost of $4$ extra DOF.

\begin{figure}[tb]
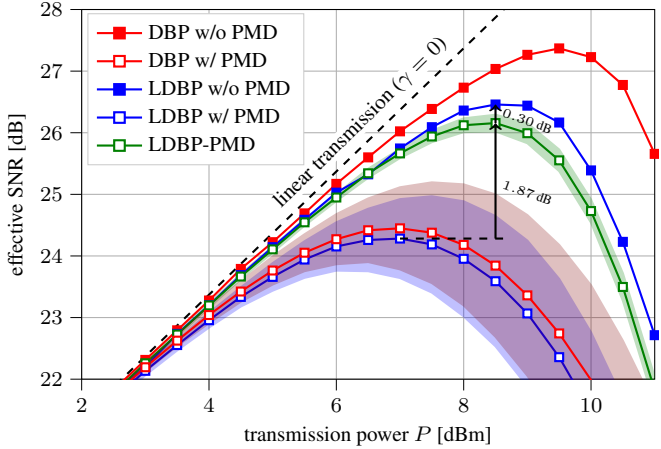

    \centering
    \includestandalone{butler8}
    \vspace{-0.25cm}
    \caption{Results for LDBP-PMD with ``free DGD filters + SU(2)$^*$ matrices'' parameterization compared to various baselines, for $40$ PMD realization. Shaded regions are standard deviations. Rotation matrices were initialized randomly.} \label{fig:LDBP-PMD_Taps5_SU2_Uniform}
\end{figure}

We consider the ``free DGD filters + SU(2)$^*$ matrices'' parameterization, as it provides a trade-off between performance and complexity in comparison to the others. 
The SNR mean and standard deviation of the considered parameterization are shown as the green curve in Fig.~\ref{fig:LDBP-PMD_Taps5_SU2_Uniform} and compared to all baselines. 
\RevB{As an upper performance bound, results for standard DBP without PMD assuming $1000$ uniform StPS (same as for the forward propagation) is also included.}
\RevB{The peak mean SNR of LDBP-PMD} is only $0.30$ dB \RevB{($1.21$ dB)} lower \RevA{than LDBP} \RevB{(DBP)} \RevA{without assuming any PMD}, and \NoRev{on average} $1.87$ dB \RevB{($1.71$ dB)} higher than LDBP \RevB{(DBP)} with an ideal frequency-domain MIMO filter appended. Additionally, it decreases the standard deviation at peak mean SNR by \RevB{at least a factor $4.5$ with respect to these last two cases}.

\RevA{When it comes to comparing the total complexity between schemes, the main difference between DBP and LDBP is that the linear steps in LDBP are trainable filters, whereas in DBP such steps are represented by fixed chromatic-dispersion filters. 
However, we have shown in previous work that the filters in LDBP can be pruned to very short lengths (as few as $3$ taps/step) \cite{Haeger2018isit, Fougstedt2018ecoc}. 
Thus, each LDBP step can be implemented efficiently in the time domain. 
When it comes to comparing LDBP+MIMO and LDBP-PMD, we only need to compare the complexity of the MIMO filter (for LDBP+MIMO) to the combined complexity of all the distributed PMD operations in LDBP-PMD.
As an example, we assume that the MIMO filter has length $16$ and is implemented as a full $4 \times 4$ filter, as proposed in \cite{Crivelli2014}. 
In this case the number of RMs and trainable parameters (DOF) is $256$. 
In terms of trainable parameters, this is comparable to the considered LDBP-PMD models, which have $41 \cdot 9 = 369$ DOF for the ``free DGD filters + SU(2)$^*$ matrices'' and $41 \cdot 5 = 205$ DOF for the ``Lagrange filters + SU(2)$^*$ matrices''. 
On the other hand, the number of RMs is $41 \cdot 36 = 1476$ and, thus, significantly larger compared to using a single MIMO filter. 
Here, we see essentially two possibilities to reduce complexity. First, while we have assumed that distributed PMD operations are performed in each step of the model, complexity could be reduced by performing such operations less frequently, e.g., only once per span, similar to \cite{Liga2018}. 
The second possibility is to reduce the per-step complexity, e.g., by reducing the length of the DGD filters or by simplifying the rotation operations. However, this may come at the price of a performance penalty, see Sec.~\ref{sec:dgd_filter_lengths} below. }

\subsection{Learning a Single Lumped MIMO-FIR Equalizer} \label{sec:results:lumpedMIMO}

\begin{figure}[tb]
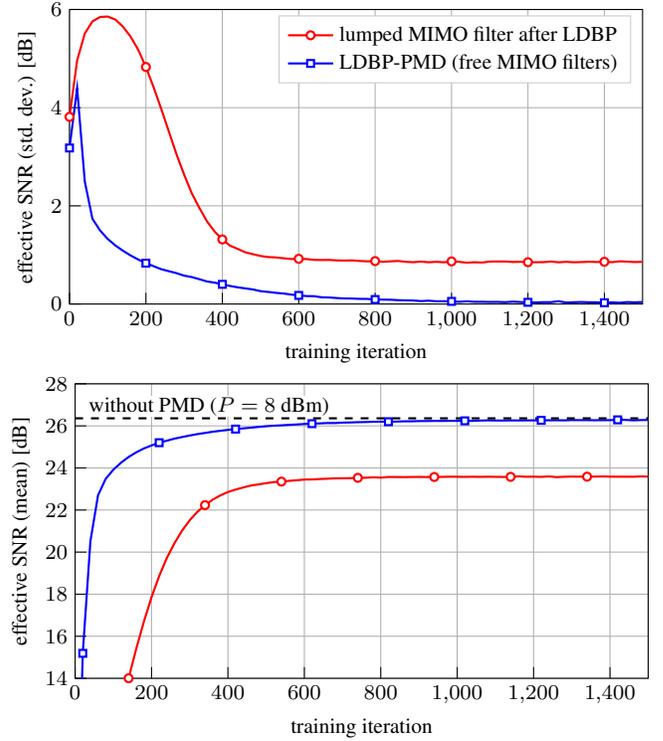
%
    \centering%
    \includestandalone{butler9top}
    \includestandalone{butler9bottom}%
    \vspace{-0.25cm}%
    \caption{Comparison of distributed and lumped MIMO-FIR filters in terms of mean effective SNR (bottom) and standard deviation (top). The distributed filters have length $5$, while the lumped filter has length $205$.}%
    \label{fig:LDBPLumpedMIMO}%
\end{figure}%

In practical systems, PMD compensation is typically performed in a lumped fashion after static equalization by optimizing the coefficients of a MIMO-FIR filter using SGD. Thus, rather than using an ideal frequency-domain MIMO filter, in this section we consider the case where the MIMO filter after LDBP is learned from data, where our main goal is to compare the convergence behavior to the distributed case. To that end, instead of learning $41$ length-$5$ MIMO-FIR filters, a single length-$5\cdot 41=205$ MIMO-FIR filter is appended to the converged LDBP model and optimized. The learning rate was kept the same as for the ``free MIMO filters'' parameterization. The resulting learning performance is shown in Fig.~\ref{fig:LDBPLumpedMIMO}. This case converges to a mean SNR of $23.59$ dB with a standard deviation of $0.86$ dB at $P = 8$ dBm transmission power. As expected, this is slightly worse than the ideal frequency-domain MIMO filter in Fig.~\ref{fig:LDBP-PMD_Taps5_SU2_Uniform}. In terms of convergence behavior, the lumped MIMO filter reaches convergence after an average of $415$ training iterations. This is only slightly faster than the $443$ iterations for LDBP-PMD with ``free MIMO filters''. This suggests that the amount of training data required for distributed PMD compensation is comparable to state-of-the-art lumped PMD compensation algorithms in practice.

\subsection{Varying the DGD Filter Length}
\label{sec:dgd_filter_lengths}

\begin{figure}[tb]
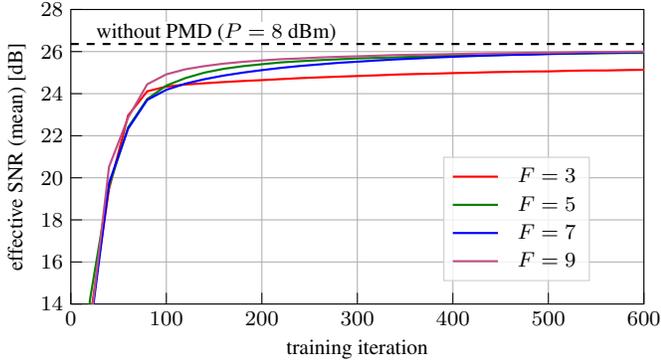

    \centering
    \includestandalone{butler10}
    \caption{Mean learning curve over all $40$ PMD realizations for the ``free DGD filters + SU(2)$^{*}$ matrices'' parameterization with random initialization for several DGD filter length $F$.} \label{fig:LDBP-PMD_TapsVary_SU2_Uniform}
\end{figure}

\RevA{In all previous scenarios, the DGD filter length was always set to $F=5$.
In this section, we study the trade-off between performance vs.~complexity when varying the DGD filter length $F$.
To that end, Fig.~\ref{fig:LDBP-PMD_TapsVary_SU2_Uniform} shows the learning curves for the ``free DGD filters + SU(2)$^*$ matrices'' parameterization assuming several DGD filter lengths $F \in \{3,5,7,9\}$. 
These results show that reducing the filter length from $F=5$ to $F=3$ results in a mean SNR penalty of around $1.0\,$dB after $600$ training iterations.
On the other hand, the complexity associated with implementing the DGD filters (which scales linearly with $F$) is reduced by $40\%$.
Fig.~\ref{fig:LDBP-PMD_TapsVary_SU2_Uniform} also shows that further increasing the filter length beyond $F=5$ to either $F=7$ or $F=9$ does not result in a noticeable improvement in terms of mean effective SNR. 
This is somewhat expected since the DGD per step is typically small. Thus, short fractional-delay filters should be sufficient for the compensation. }
\RevB{In that regard, it is also interesting to mention the possibility of compensating DGD with pure delay elements instead of fractional-delay filters. 
This would significantly reduce the computational complexity by avoiding any multiplications or additions.
For example, it has been shown that the walk-off delay in subband-based linear and nonlinear equalizers can (at least partially) be performed with delay elements \cite{Ho2009, Haeger2018ecoc}.
However, it is not immediately clear how this approach could be applied to PMD compensation. 
Indeed, the DGDs are randomly distributed, whereas \cite{Ho2009, Haeger2018ecoc} exploit the fact that the walk-off delay is a deterministic function of the chosen sampling rate.}

\subsection{Instantaneous PMD Changes} \label{sec:results:adaptive}

\begin{figure}
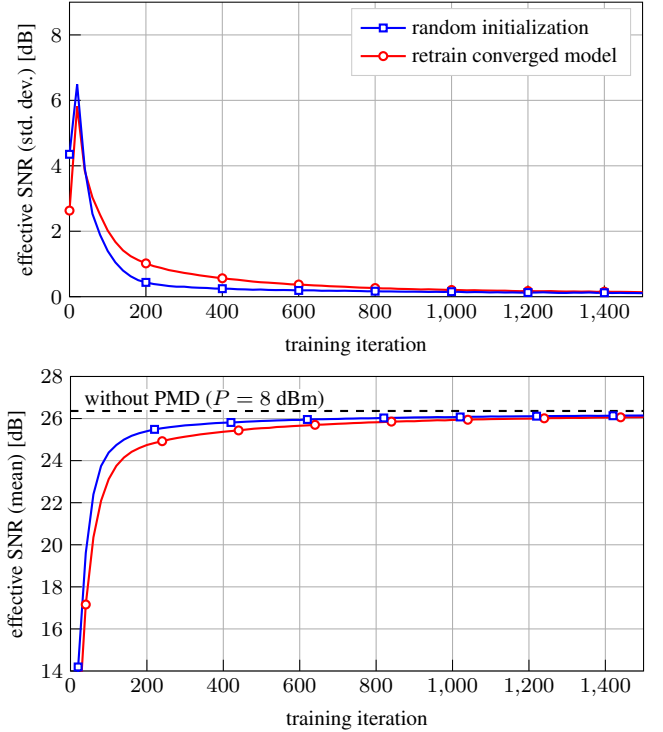

    \centering
    \includestandalone{butler11top}
    \includestandalone{butler11bottom}
    \vspace{-0.25cm}
    \caption{Results for instantaneous PMD changes in terms of mean effective SNR (bottom) and standard deviation (top), where the model is based on the ``free DGD filters + SU(2)$^*$ matrices'' parameterization. }
    \label{fig:adaptive}
\end{figure}

So far, all considered links comprised a static PMD scenario. However, in practice, PMD may drift over time, calling for an adaptive scheme. To conclude this paper, we consider a scenario where the PMD realization changes instantaneously and is uncorrelated with its prior state. To that end, each converged LDBP-PMD model with the ``free DGD filters + SU(2)$^*$ matrices'' parameterization and random initialization is retrained for $1,500$ additional iterations on a different PMD realization, randomly chosen from the $39$ remaining realizations that were not used for the initial training. The resulting learning performance is shown in Fig.~\ref{fig:adaptive}. In this case, the model converges slower compared to the case where it is randomly initialized, taking an average of $657$ iterations instead of the original $428$. However, it does converge to a mean SNR of $26.07$ dB with a standard deviation of $0.14$, which is only $0.05$ dB lower and a factor $1.1$ higher than the original case. 
\RevA{While these results suggest that training again ``from scratch'' may be better compared to retraining the already converged model, it is not clear if the same conclusion also holds if an abrupt PMD change occurs only ``locally'', i.e., over a short propagation distance. 
A closer investigation of this question is left for future work.}

\section{Conclusions}\label{sc:conc}

We have proposed a multi-layer machine-learning model for DP systems based on parameterizing the SSFM for the Manakov-PMD equation. As an application, we have considered receiver-side DBP and distributed PMD compensation, where we investigated several parameterization choices and compared them in terms of performance, complexity, and convergence behavior. In general, parameterizations with more DOF can attain better performance but may also require more training iterations to convergence. A good compromise between performance and complexity is obtained by freely optimizing short real-valued DGD filters combined with SU(2)-rotating matrices without the unitary constraint. We demonstrated that our model provides performance close to the PMD-free case, without any prior knowledge about the PMD realizations along the link or the total accumulated PMD.

\ifthenelse{\isthesis=1}{%
\section{Acknowledgements}
This Master Thesis was supervised by Alex Alvarado, in collaboration with Christian H\"ager and Gabriele Liga. Together with Frans Willems, they composed the defense committee. I would like to express my sincere gratitude towards them for presenting me the opportunity to carry out this research, guiding me every step along the way, providing me with constructive feedback on my performance, and teaching me everything I know about communication theory.

I also wish to thank Vinícius Oliari, Astrid Barreiro and Kaiquan Wu for aiding me with my theoretical questions, and bearing with me until I fully understood each answer.

I would like to extend my gratitude to all other professors, assistant professors, and doctoral candidates who taught me the skills necessary during my studies to complete this project. I am especially grateful towards Chigo Okonkwo, Hugo de Waardt, and Nicola Calabretta for teaching me about the physics of optical fiber links, to Ruud van Sloun, Fons van der Sommen, Rik Vullings, Joost van der Putten, and Egor Bondarau for providing me with a strong knowledge base on the topic of Deep Learning, and to Simona Turco, Firat Tigrek, Wim van Houtum, and Yunus Gültekin for reinforcing my understanding of communication theory.

I gratefully acknowledge the role Henry Pfister played in the fundamental research of LDBP. Without his contributions, this work would not have been possible.

Furthermore, I would like to express my gratitude towards everyone at ICTLAB, who welcomed me into their research group and presented me with many opportunities for my career and personal development.

Finally, I thank my parents Hans and Petra, my brother Frank, and my grandparents, for their support at home. Thanks also to my dear friends, without whom the last half year would not have been nearly as entertaining.
}{}

\balance


\begin{thebibliography}{10}
\providecommand{\url}[1]{#1}
\csname url@samestyle\endcsname
\providecommand{\newblock}{\relax}
\providecommand{\bibinfo}[2]{#2}
\providecommand{\BIBentrySTDinterwordspacing}{\spaceskip=0pt\relax}
\providecommand{\BIBentryALTinterwordstretchfactor}{4}
\providecommand{\BIBentryALTinterwordspacing}{\spaceskip=\fontdimen2\font plus
\BIBentryALTinterwordstretchfactor\fontdimen3\font minus
  \fontdimen4\font\relax}
\providecommand{\BIBforeignlanguage}[2]{{%
\expandafter\ifx\csname l@#1\endcsname\relax
\typeout{** WARNING: IEEEtran.bst: No hyphenation pattern has been}%
\typeout{** loaded for the language `#1'. Using the pattern for}%
\typeout{** the default language instead.}%
\else
\language=\csname l@#1\endcsname
\fi
#2}}
\providecommand{\BIBdecl}{\relax}
\BIBdecl

\bibitem{Monroe2016}
D.~Monroe, ``Optical fibers getting full,'' \emph{Commun. ACM}, vol.~59,
  no.~10, p. 10–12, Sep. 2016.

\bibitem{Stolen1973}
R.~Stolen and A.~Ashkin, ``Optical {Kerr} effect in glass waveguide,''
  \emph{Appl. Phys. Lett.}, vol.~22, pp. 294--296, Mar. 1973.

\bibitem{Kao1966}
K.~C. Kao and G.~A. Hockham, ``Dielectric-fibre surface waveguides for optical
  frequencies,'' \emph{IEE Proc. J. Optoelectronics}, vol. 113, no.~7, pp.
  191--198, Jun. 1986.

\bibitem{Nagel1989}
S.~R. Nagel, ``Optical fibre-the expanding medium,'' \emph{IEEE Circuits and
  Devices Magazine}, vol.~5, no.~2, pp. 36--45, Mar. 1989.

\bibitem{Ip2010a}
E.~Ip, ``{Nonlinear compensation using backpropagation for
  polarization-multiplexed transmission},'' \emph{J. Lightw. Technol.},
  vol.~28, no.~6, pp. 939--951, Mar. 2010.

\bibitem{Menyuk2004b}
C.~R. Manyuk, ``Interaction of nonlinearity and polarization mode dispersion,''
  in \emph{Polarization Mode Dispersion}, A.~Galtarossa and C.~R. Menyuk,
  Eds.\hskip 1em plus 0.5em minus 0.4em\relax New York: Springer, 2004, pp.
  126--132.

\bibitem{Shen2011}
T.~S.~R. Shen and A.~P.~T. Lau, ``Fiber nonlinearity compensation using extreme
  learning machine for {DSP}-based coherent communication systems,'' in
  \emph{Proc. Optoelectronics and Communications Conf. (OECC)}, Kaohsiung,
  Taiwan, 2011, pp. 816--817.

\bibitem{Jarajreh2015}
A.~M. Jarajreh, E.~Giacoumidis, I.~Aldaya, S.~T. Le, A.~Tsokanos,
  Z.~Ghassemlooy, and N.~J. Doran, ``{Artificial Neural Network Nonlinear
  Equalizer for Coherent Optical {OFDM}},'' \emph{IEEE Photon. Technol. Lett.},
  vol.~27, no.~4, pp. 387--390, Feb. 2015.

\bibitem{Estaran2016}
J.~Estar{\'{a}}n, R.~Rios-M{\"{u}}ller, M.~A. Mestre, F.~Jorge, H.~Mardoyan,
  A.~Konczykowska, J.~Y. Dupuy, and S.~Bigo, ``{Artificial Neural Networks for
  Linear and Non-Linear Impairment Mitigation in High-Baudrate {IM}/{DD}
  Systems},'' in \emph{Proc. European Conf. Optical Communication (ECOC)},
  D{\"{u}}sseldorf, Germany, 2016, pp. 106--108.

\bibitem{OShea2017}
T.~O'Shea and J.~Hoydis, ``{An Introduction to Deep Learning for the Physical
  Layer},'' \emph{IEEE Trans. Cogn. Commun. Netw.}, vol.~3, no.~4, pp.
  563--575, Dec. 2017.

\bibitem{Shen2018ecoc}
S.~Li, C.~H{\"{a}}ger, N.~Garcia, and H.~Wymeersch, ``Achievable information
  rates for nonlinear fiber communication via end-to-end autoencoder
  learning,'' in \emph{Proc. European Conf. Optical Communication (ECOC)},
  2018, pp. 1--3.

\bibitem{Jones2018}
R.~T. Jones, T.~A. Eriksson, M.~P. Yankov, and D.~Zibar, ``{Deep Learning of
  Geometric Constellation Shaping including Fiber Nonlinearities},'' in
  \emph{Proc. European Conf. Optical Communication (ECOC)}, Rome, Italy, 2018,
  pp. 1--3.

\bibitem{Karanov2018}
B.~{Karanov}, M.~{Chagnon}, F.~{Thouin}, T.~A. {Eriksson}, H.~{Bülow},
  D.~{Lavery}, P.~{Bayvel}, and L.~{Schmalen}, ``End-to-end deep learning of
  optical fiber communications,'' \emph{J. Lightw. Technol.}, vol.~36, no.~20,
  pp. 4843--4855, Oct. 2018.

\bibitem{Khan2019}
F.~N. Khan, Q.~Fan, C.~Lu, and A.~P.~T. Lau, ``{An optical communication's
  perspective on machine learning and its applications},'' \emph{J. Lightw.
  Technol.}, vol.~37, no.~2, pp. 493--516, Jan. 2019.

\bibitem{Haeger2018ofc}
C.~H{\"{a}}ger and H.~D. Pfister, ``{Nonlinear Interference Mitigation via Deep
  Neural Networks},'' in \emph{Proc. Optical Fiber Communication Conf. (OFC)},
  San Diego, CA, 2018, pp. 1--3.

\bibitem{Haeger2018isit}
C.~H{\"{a}}ger and H.~D. Pfister, ``{Deep Learning of the Nonlinear {S}chr{\"{o}}dinger Equation in
  Fiber-Optic Communications},'' in \emph{Proc. IEEE Int. Symp. Information
  Theory (ISIT)}, 2018, pp. 1590--1954.

\bibitem{Lian2018itw}
M.~Lian, C.~H{\"{a}}ger, and H.~D. Pfister, ``{What Can Machine Learning Teach
  Us about Communications?}'' in \emph{Proc. IEEE Information Theory Workshop
  (ITW)}, Guangzhou, China, 2018, pp. 1--5.

\bibitem{Haeger2019ecoc}
C.~H{\"{a}}ger, H.~D. Pfister, R.~M. B{\"{u}}tler, G.~Liga, and A.~Alvarado,
  ``{Revisiting Multi-Step Nonlinearity Compensation with Machine Learning},''
  in \emph{Proc. European Conf. Optical Communication (ECOC)}, Dublin, Ireland,
  2019, pp. 1--4.

\bibitem{Oliari2020}
V.~Oliari, S.~Goossens, C.~H{\"{a}}ger, G.~Liga, R.~M. B{\"{u}}tler, M.~van~den
  Hout, S.~van~der Heide, H.~D. Pfister, C.~Okonkwo, and A.~Alvarado,
  ``{Revisiting Efficient Multi-Step Nonlinearity Compensation with Machine
  Learning: An Experimental Demonstration},'' \emph{J. Lightw. Technol.},
  vol.~38, no.~12, pp. 3114--3124, Jun. 2020.

\bibitem{Haeger2020ofc}
C.~H\"{a}ger, H.~D. Pfister, R.~M. B\"{u}tler, G.~Liga, and A.~Alvarado,
  ``Model-based machine learning for joint digital backpropagation and pmd
  compensation,'' in \emph{Proc. Optical Fiber Communication Conf. (OFC)}, Los
  Angeles, CA, 2020, pp. 1--3.

\bibitem{Nelson2005}
L.~E. Nelson and R.~M. Jopson, ``Introduction to polarization mode dispersion
  in optical systems,'' \emph{J. Optic. Comm. Rep.}, vol.~1, no.~4, pp.
  312--344, Dec. 2004.

\bibitem{Goroshko2016}
K.~Goroshko, H.~Louchet, and A.~Richter, ``{Overcoming performance limitations
  of digital back propagation due to polarization mode dispersion},'' in
  \emph{Proc. Int. Conf. Transparent Optical Networks (ICTON)}, Trento, Italy,
  2016, pp. 1--4.

\bibitem{Czegledi2016a}
C.~B. Czegledi, G.~Liga, D.~Lavery, M.~Karlsson, E.~Agrell, S.~J. Savory, and
  P.~Bayvel, ``{Polarization-Mode Dispersion Aware Digital Backpropagation},''
  in \emph{Proc. European Conf. Optical Communication (ECOC)},
  D{\"{u}}sseldorf, Germany, 2016, pp. 1--3.

\bibitem{Czegledi2017}
C.~B. Czegledi, G.~Liga, D.~Lavery, M.~Karlsson, E.~Agrell, S.~J.
  Savory, and P.~Bayvel, ``{Digital backpropagation accounting for
  polarization-mode dispersion},'' \emph{Opt. Express}, vol.~25, no.~3, pp.
  1903--1915, Feb. 2017.

\bibitem{Liga2018}
G.~Liga, C.~Czegledi, and P.~Bayvel, ``A {PMD}-adaptive {DBP} receiver based on
  {SNR} optimization,'' in \emph{Proc. Optical Fiber Communication Conf.
  (OFC)}, San Diego, CA, 2018, pp. 1--3.

\bibitem{Fougstedt2017}
C.~Fougstedt, M.~Mazur, L.~Svensson, H.~Eliasson, M.~Karlsson, and
  P.~Larsson-Edefors, ``{Time-Domain Digital Back Propagation: Algorithm and
  Finite-Precision Implementation Aspects},'' in \emph{Proc. Optical Fiber
  Communication Conf. (OFC)}, Los Angeles, CA, 2017, pp. 1--3.

\bibitem{Fougstedt2018ecoc}
C.~Fougstedt, C.~H{\"{a}}ger, L.~Svensson, H.~D. Pfister, and
  P.~Larsson-Edefors, ``{ASIC Implementation of Time-Domain Digital
  Backpropagation with Deep-Learned Chromatic Dispersion Filters},'' in
  \emph{Proc. European Conf. Optical Communication (ECOC)}, Rome, Italy, 2018,
  pp. 1--4.

\bibitem{Goodfellow2016}
I.~Goodfellow, Y.~Bengio, and A.~Courville, \emph{Deep Learning}, 1st~ed.\hskip
  1em plus 0.5em minus 0.4em\relax MIT Press, 2016.

\bibitem{Nielsen2015}
M.~A. Nielsen, \emph{Neural Networks and Deep Learning}, 1st~ed.\hskip 1em plus
  0.5em minus 0.4em\relax Determination Press, 2015.

\bibitem{Kingma2014a}
D.~P. Kingma and J.~Ba, ``{Adam: A Method for Stochastic Optimization},'' in
  \emph{Proc. Int. Conf. Learning Representations (ICLR)}, San Diego, CA, 2015,
  pp. 1--15.

\bibitem{Savory2008}
S.~J. Savory, ``{Digital filters for coherent optical receivers},'' \emph{Opt.
  Express}, vol.~16, no.~2, p. 804–817, Jan. 2008.

\bibitem{Wai1991}
P.~K.~A. Wai, C.~R. Menyuk, and H.~H. Chen, ``{Stability of solitons in
  randomly varying birefringent fibers},'' \emph{Opt. Lett.}, vol.~16, no.~16,
  p. 1231, Aug. 1991.

\bibitem{Agrawal2019}
G.~P. Agrawal, \emph{Nonlinear Fiber Optics}, 6th~ed.\hskip 1em plus 0.5em
  minus 0.4em\relax Academic Press, 2019.

\bibitem{Ellis2019}
A.~Ellis and M.~Sorokina, \emph{Optical Communication Systems}, 1st~ed.\hskip
  1em plus 0.5em minus 0.4em\relax New York: Jenny Stanford Publishing, 2019.

\bibitem{Essiambre2005}
R.-J. Essiambre and P.~J. Winzer, ``{Fibre Nonlinearities in Electronically
  Pre-Distorted Transmission},'' in \emph{Proc. European Conf. Optical
  Communication (ECOC)}, Glasgow, UK, 2005, pp. 1--3.

\bibitem{Roberts2006}
K.~Roberts, C.~Li, L.~Strawczynski, M.~O'Sullivan, and I.~Hardcastle,
  ``{Electronic precompensation of optical nonlinearity},'' \emph{IEEE Photon.
  Technol. Lett.}, vol.~18, no.~2, pp. 403--405, Jan. 2006.

\bibitem{Lavery2016a}
D.~Lavery, D.~Ives, G.~Liga, A.~Alvarado, S.~J. Savory, and P.~Bayvel, ``{The
  Benefit of Split Nonlinearity Compensation for Single-Channel Optical Fiber
  Communications},'' \emph{IEEE Photon. Technol. Lett.}, vol.~28, no.~17, pp.
  1803--1806, Sep. 2016.

\bibitem{Kootsookos1996}
P.~J. Kootsookos and R.~C. Williamson, ``{FIR} approximation of fractional
  sample delay systems,'' \emph{IEEE Trans. Circuits Syst. II: Analog and
  Digital Signal Proc.}, vol.~43, no.~3, pp. 269--271, Mar. 1996.

\bibitem{Zhang2013}
J.~Zhang, X.~Li, and Z.~Dong, ``Digital nonlinear compensation based on the
  modified logarithmic step size,'' \emph{J. Lightw. Technol.}, vol.~31,
  no.~22, pp. 3546--3555, Nov. 2013.

\bibitem{Crivelli2014}
D.~E. Crivelli, M.~R. Hueda, H.~S. Carrer, M.~{Del Barco}, R.~R. L{\'{o}}pez,
  P.~Gianni, J.~Finochietto, N.~Swenson, P.~Voois, and O.~E. Agazzi,
  ``{Architecture of a single-chip 50 {Gb/s} {DP}-{QPSK/BPSK} transceiver with
  electronic dispersion compensation for coherent optical channels},''
  \emph{IEEE Trans. Circuits Syst. I: Reg. Papers}, vol.~61, no.~4, pp.
  1012--1025, Apr. 2014.

\bibitem{Ho2009}
K.-P. Ho, ``{Subband equaliser for chromatic dispersion of optical fibre},''
  \emph{Electronics Lett.}, vol.~45, no.~24, pp. 1224--1226, Nov. 2009.

\bibitem{Haeger2018ecoc}
C.~H{\"{a}}ger and H.~D. Pfister, ``{Wideband Time-Domain Digital
  Backpropagation via Subband Processing and Deep Learning},'' in \emph{Proc.
  European Conf. Optical Communication (ECOC)}, Rome, Italy, 2018, pp. 1--3.

\end{thebibliography}
\end{document}